\documentstyle{mn}

\input{epsf}
\begin{document}

\title[Large Scale Fluctuations in Galaxy Distribution]
{Large scale fluctuations in the distribution of galaxies}

\author[C.M. Baugh]
{C.M. Baugh 
 \\
Department of Physics, Science Laboratories, South Road, Durham DH1 3LE}

\maketitle

\def\mpc {h^{-1} {\rm{Mpc}}}
\def\and  {\it {et al.} \rm}
\def\rmd {\rm d}

\begin{abstract}
The null hypothesis that the three dimensional power spectrum measured 
from the APM Survey is consistent with the one dimensional power 
spectrum measured from the pencil beams surveys of Broadhurst \and  and 
of Szalay \and is tested.
The external estimates of the mean power that we make are
sensitive to details of the model for the survey geometry and
to the assumed level of the distortion of the pattern of
galaxy clustering caused by peculiar motions.
We find that the measured 3D clustering of galaxies can account
for the presence of peaks in the one dimensional power spectrum,
but is less successful in recovering the detailed appearence of
the observations.
We find no strong evidence for any additional large scale structure in the 
deep pencil beams beyond that recovered from the APM Survey.
We conclude therefore that it is unlikey that large scale structure can be
responsible for the steep local number counts of bright galaxies.
\end{abstract}


\section{Introduction}

The claim by Broadhurst \and (1990, hereafter BEKS) 
that the universe contains 
significant structure on scales larger than $\sim 100 \mpc$ 
has provoked much controversy.
Based upon the analysis of several deep, narrow angle or pencil beam 
redshift surveys of galaxies in the directions of the Galactic Poles,
BEKS noted a striking periodicity in the pair counts of the 
galaxies.
This periodicity is revealed as a spike in the one dimensional power 
spectrum of the radial galaxy distribution at 
$ k  \sim 0.05 h {\rm Mpc}^{-1}$, corresponding to a wavelength of 
$ \lambda \sim 128 \mpc$.
Using estimates of the mean power from the observed power 
spectrum (internal) and using a model for galaxy clustering 
combined with a window function for the survey (external), BEKS 
attatched a high significance to the peak.
The adopted null hypothesis, namely that the two point correlation 
function of galaxies on small scales has a power law form 
$ \xi(r) = (r/r_{0})^{\gamma} $ (Davis \& Peebles 1983), with 
a truncation at $r = 30 \mpc$ and random correlations on 
larger scales, was ruled out.

In a comprehensive paper, Kaiser \& Peacock (1991 - KP) argued that 
the mean power had been underestimated, which would reduce the 
significance of the power spectrum peak.
Using a simple Poisson clumps model, KP demonstrated that the power 
spectrum at higher wavenumbers could be suppressed by peculiar velocities and 
redshift errors, leading to a low internal estimate of the mean 
power.
Furthermore, with more detailed modelling of the BEKS survey geometry 
and radial selection, KP showed that the external estimate of the 
power could be a factor of two larger than the figure obtained by BEKS.
Note that KP employ a different null hypothesis, in that the observed 
small scale clustering of galaxies is extended to all scales 
using the same power 
law.

Szalay \and (1993) remark that assessments of the significance of the 
peak in the BEKS power spectrum based upon external estimates of the 
mean power are unreliable.
Such estimates are sensitive to assumptions 
made about the clustering of galaxies on large scales and about the 
survey geometry.
This point is re-enforced by Luo \& Vishniac (1993), who using the same 
model for galaxy clustering 
but a slightly different window geometry and selection 
function to KP, obtain a value for the mean power that is 
intermediate between 
that of Szalay \and (1993) and KP.

The original BEKS data has now been supplemented by additional pencil 
beams giving a wider angular coverage (Szalay \and 1993, 
Broadhurst \and 1995). These authors claim 
that the aliasing of small scale power to large scales is reduced 
for the extended survey geometry, which has a narrower window function in 
$k$-space.
In addition, there are new claims for excess power on
scales of $\lambda \sim 100 h^{-1} {\rm Mpc}$, compared with
standard models of structure formation such as the Cold
Dark Matter scenario and its variants, from power spectrum
analysis of the slices of the Las Campanas Redshift Survey
(Landy \and 1996), though the longest baseline of this survey
is shorter than that of the BEKS data.

Measurements of the three dimensional power spectrum 
on scales $ \lambda > 100 \mpc $ are now available from 
large redshift surveys ({\it e.g.} Feldman, Kaiser \& Peacock 1994, 
Tadros \& Efstathiou 1996) and from the deprojection of the 
clustering information in angular catalogues such as the APM Survey  
(Peacock 1991, Baugh \& Efstathiou 1993, 1994).

In this paper we shall ask the question is the three dimensional 
power spectrum measured in real space from the APM Survey 
(Baugh \& Efstathiou 1993, 1994) consistent with the one 
dimensional power spectrum of the BEKS survey and the new data 
reported by Szalay \and (1993) and Broadhurst \and (1995).
The models adopted for the survey geometry of the observations
are outlined in Section \ref{s:pen}, with the effects of including
redshift space distortions discussed in Section \ref{s:red}.
The significance of the observed peaks are assessed by comparison
with our external estimates of the mean power in
Section \ref{s:peak}.

\section{Pencil beam window functions}
\label{s:pen}

The projected one dimensional power spectrum is a convolution of 
the three dimensional power spectrum with the Fourier transform of 
the window function of the survey:

\begin{equation}
P_{1D}(k) = \frac{1}{2 \pi} \int_{k}^{\infty} {\rm d} y \, y 
P_{3D} (y)  I (k,y),
\label{eq:p1d}
\end{equation}
where $I(k,y)$ depends upon the particular model adopted for the 
radial and angular selection of the survey.
We shall discuss three survey geometries:
\begin{itemize}

\item[{\bf A:}] Uniform cylinder: a disk of 
      fixed comoving radius $R$, neglecting 
      the radial selection function, with cylinder length $L$ given by 
      the highest redshift galaxies observed. 
      Following BEKS, we use a cylinder with radius $R= 3\mpc$ and 
      length $L=2034 \mpc$.	

\item[{\bf B:}] A conical window, with fixed opening angle and a model for 
      the survey selection function. In order to reproduce the 
       deep NGP and SGP pencil beams of BEKS, we use a conical beam 
      with opening angle $10$ arcminutes and a magnitude limit of 
      $b_{J} \le 22.5 $.

\item[{\bf C:}] Multiple conical windows 
      distributed at random within some larger solid  
      angle. We choose 21 beams with opening angle $ 15 $ arcminutes 
      distributed at random within a cone of opening angle $5 ^{\circ}$, 
      with galaxies in the magnitude range $ 17 \le b_{J} \le 20.5$, as 
      described by Szalay \and (1993).

\end{itemize}

For a uniform cylinder, to second order in 
$k$ the window function takes the form 
(following BEKS, KP equation (3.10))

\begin{equation}
I(k,y) = \frac{1}{L} \exp ( - \left[\sqrt(y^{2}-k^{2}) R\right]^{2} /4 ).
\end{equation}

\begin{figure}
\centering
\centerline
{\epsfxsize=8.5truecm \epsfysize=8.5truecm 
\epsfbox{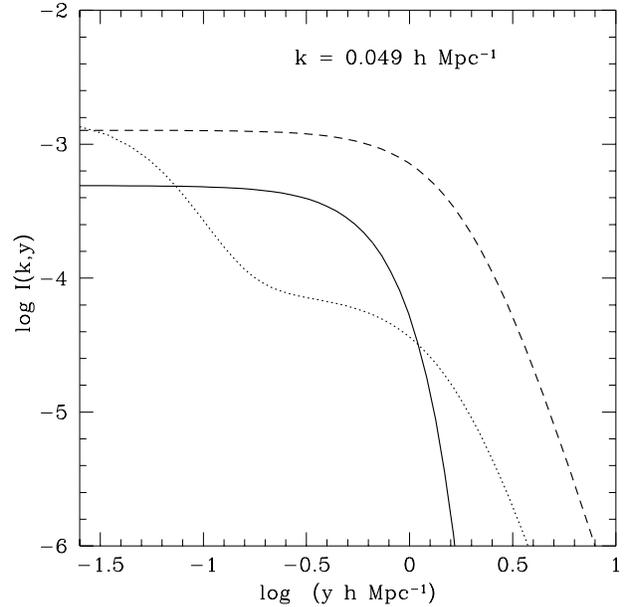}}
\caption[junk]{The Fourier transform of the window functions for 
the survey geometries considered here.
The solid line shows a uniform cylinder, the dashed line shows  
a conical window with a radial selection function and the dotted 
line shows the window function for multiple pencil beams.
}
\label{fig:wk}
\end{figure}

To calculate the window function for a conical survey including the 
effects of a radial selection function, we follow the derivation given in 
Appendix A1 of KP, substituting the redshift distribution of 
galaxies per steradian $ {\rm d} {\cal N} / {\rm d} z$ used by 
Baugh \& Efstathiou (1993): 

\begin{equation}
{\rm d} {\cal N}/ {\rm d} z = \frac{3 {\cal N}}
{2 z_{c}^{3}} \exp ( - (z/z_{c})^{1.5}),
\label{eq:dndz}
\end{equation}
where the median redshift is given by
\begin{eqnarray}
z_m(b_{J}) &=& 1.412 z_{c} \\
           &=& 0.016 (b_{J} - 17)^{1.5} + 0.046
\end{eqnarray}
This parametric form for the redshift distribution 
was chosen to fit both the Stromlo-APM 
survey (Loveday \and 1992) 
and the fainter surveys of Broadhurst \and (1988) and 
Colless \and (1990, 1993) 
({\it cf} Figure 1 of Baugh \& 
Efstathiou 1993).

With this model for the selection function, the window function is 
given by:

\begin{equation}
I(k,y) = \frac{\int_{0}^{\infty} {\rm d} x \left ( {\rm d} {\cal N} /
{\rm d} x \right)^{2} W^{2} (\theta x \sqrt(y^{2} - k^{2}))}
{\left[ \int_{0}^{\infty} {\rm d} z 
\left( {\rm d} {\cal N}/ {\rm d} z \right)\right]^{2}}, 
\end{equation}
where the angular window function is normalised to unity.
For a single beam, the Fourier transform of the angular selection 
function is given by

\begin{equation}
W(k \theta) = \exp( - (k \theta)^{2}/8 ),
\label{eq:wang}
\end{equation}
whilst for $N$ beams each of opening angle $\theta_{1}$ placed at 
random within a solid angle of opening angle $\theta_{2}$, the angular 
selection is given by (equation 4.5 KP):

\begin{equation}
W(k) = W (k \theta_{1}) ( W (k \theta_{2}) + 1/N), 
\end{equation} 
with $W (k \theta)$ given by equation \ref{eq:wang}.

The form of the window function Fourier transforms given above 
are plotted in Figure \ref{fig:wk}.
This Figure shows the large effect that the adopted survey 
geometry can have upon the deduced mean power; increasing the 
number of pencil beams damps the aliasing of power from small 
scales.

\section{Redshift Space Distortions}
\label{s:red}

\begin{figure}
\centering
\centerline
{\epsfxsize=8.5truecm \epsfysize=8.5truecm 
\epsfbox{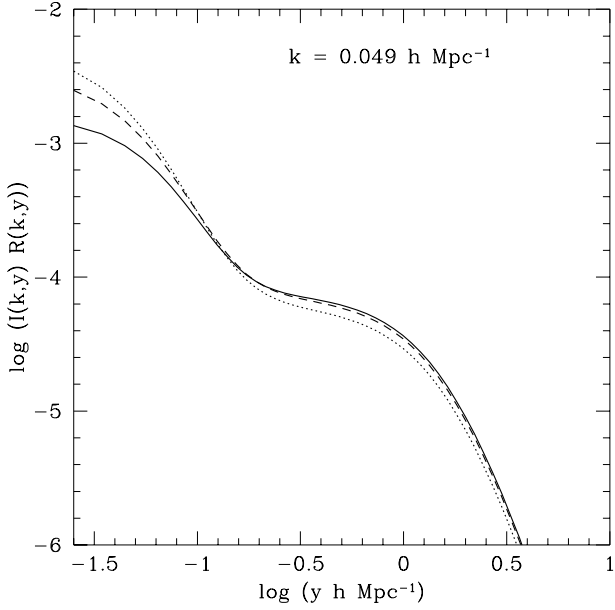}}
\caption[junk]{
The Fourier transform of the window functions for 
the Case C, multiple beams.
The solid line shows the window function in real space.
The dotted line shows the window function when redshift 
distortions characterised by $\beta=1.0$ and $\sigma_{v} = 1000 {\rm kms}^{-1}$ 
are included; the dashed line shows the window function for $\beta=0.5$, 
$\sigma_{v} = 500 {\rm kms}^{-1}$ 
}
\label{fig:wks}
\end{figure}

The pattern of galaxy clustering measured in redshift 
space is distorted by the 
peculiar motions of galaxies.
Coherent bulk flows lead to a boost in the measured power in three dimensions on 
large scales (Kaiser 1987), whilst the virialised motions of galaxies in 
groups and clusters results in a reduction in the power on small scales. 
Hence, in order to account properly for the effects of peculiar velocities when 
calculating the projected 1D power, it is necessary to model both the small and large 
scale phenomena.

A simple approach is to assume that galaxy velocities are given by linear 
theory on large scales plus an uncorrelated, random velocity dispersion which 
dominates on small scales ({\it e.g.} Peacock \& Dodds 1994; 
Cole, Fisher \& Weinberg 1995).
The redshift space power spectrum $P_{S}(k,\mu_)$ is then given in terms of 
the real space power spectrum $P_{R}(k)$ by 

\begin{equation}
P_{S} (k, \mu) = P_{R} (k) (1 + \beta \mu^{2})^{2}/ (1 + (k \sigma_{v} \mu/H_{0})^{2}/2)^{2}
\end{equation}
where $\beta = \Omega^{0.6}/b$, $\Omega$ is the density of the 
universe in units 
of the critical density, $b$ is the bias between fluctuations in the 
galaxy distribution and the underlying 
mass distribution, $\sigma_{v}$ is the 1D velocity 
dispersion and $\mu = {\bf k} \cdot {\bf \hat{z}}/|{\bf k}|$ is the cosine of the angle between 
the line of sight and the wavevector.
We have assumed that the small scale velocities follow an exponential distribution 
as indicated by N-body simulations (Park {\it et al} 1994). 

Extending the results of Section 4.3 of KP the projected power including the 
effects of redshift space distortions on small and large scales is 
given by:

\begin{equation}
P_{1D}(k) = \frac{1}{2 \pi} \int_{k}^{\infty} {\rm d} y \, y 
P_{3D} (y)  I (k,y) R(k,y),
\label{eq:p1ds}
\end{equation}
where 
\begin{equation}
R(k,y) = (1 + \beta (k/y)^{2})^{2} / ( 1 + (\sigma_{v} k /H_{0})^{2}/2)^{2}
\end{equation}

The effects of including redshift space distortions on the shape of the 
window function in Case C, multiple pencil beams, is shown in Figure \ref{fig:wks}.
Estimates of the parameter $\beta$ show a large spread in values.
Determinations using the Stromlo/APM and
APM Surveys by Loveday \and (1995) and
Baugh (1996) find $\beta \sim 0.5$; a careful
analysis of the errors by Tadros \& Efstathiou (1996) shows however
that values as large as $ \beta \sim 1 $ cannot be ruled out by the
present generation of redshift surveys.
Tadros \& Efstathiou find one dimensional velocity dispersions in the
range $\sigma_{v} = 300-700 {\rm kms}^{-1}$ in N-body simulations of 
the standard Cold Dark
Matter model and its variants.

\section{Peak significance}
\label{s:peak}

\begin{figure*}
\centering
\centerline
{\epsfxsize=14.truecm \epsfysize=14.truecm 
\epsfbox{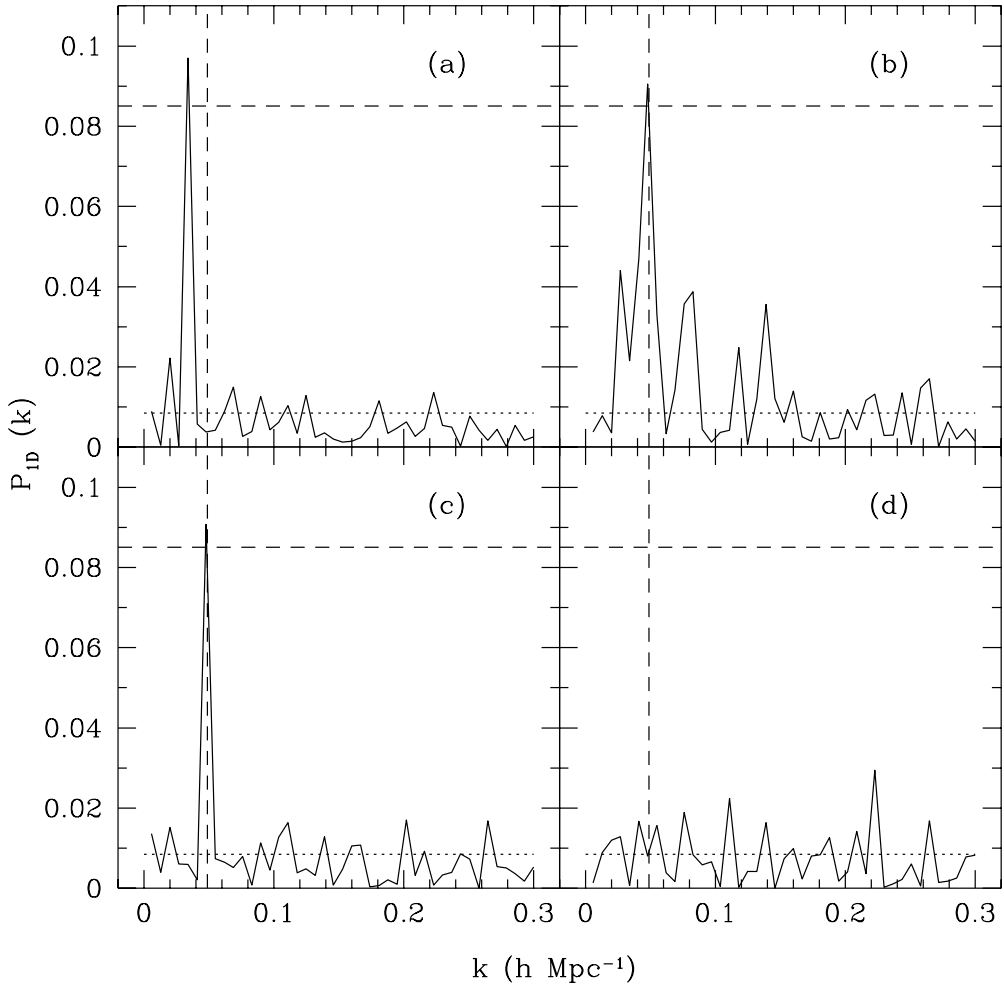}}
\caption
{
Gaussian realisations of the 1D power spectrum 
predicted by the 3D power spectrum of the APM survey after 
convolution with the survey window function of Szalay \and (1993).
The vertical dotted line shows the wavenumber of the spike in the 
power spectrum reported by BEKS and Szalay \and (1993).
The lower dashed line shows the mean power level at $k = 0.049 h {\rm 
Mpc}^{-1}$.
The upper dashed line shows the peak amplitude reported by 
Broadhurst \and (1995).
Two realisations are shown (a) and (c) that contain a sharp peak with 
no other peak higher than $P \sim 0.02$; such spectra occur much 
less often than the type shown in (b), where there are peaks several times 
higher than the mean power at other wavenumbers.
Panel (d) shows and example of a spectrum without any high peaks. 
}
\label{fig:pk}
\end{figure*}

\begin{table}
\begin{center}
\caption[dummy]{Mean power for survey geometries in real space}
\label{tab:mpr}
\begin{tabular}{ccccc}
\hline
\hline  
     &Survey             & Mean power          &  ratio           \\ 
     &geometry           & $\pm 1 \sigma $     &  $P_{peak}/<P>$  \\ \hline
  (A)&                   & $0.014 \pm 0.002$   &  15.2            \\ \hline
  (B)& $b_{J} \le 21.5$  & $0.065 \pm 0.009$   &   3.3            \\
     & $b_{J} \le 22.5$  & $0.047 \pm 0.007$   &   4.5            \\ \hline
  (C)& $\alpha=0$        & $0.0085 \pm 0.0016$ &  10.0            \\ 
     & $\alpha=2$        & $0.0106 \pm 0.0020$ &   8.0            \\ \hline
\hline
\end{tabular}
\end{center}
\end{table}

\begin{table}
\begin{center}
\caption[dummy]{Mean power for geometry C in redshift space}
\label{tab:mps}
\begin{tabular}{ccccc}
\hline
\hline  
$\beta$          &  $\sigma_{v} $   & Mean power          &  ratio           \\ 
                 &  (kms$^{-1}$)    & $\pm 1 \sigma $     &  $P_{peak}/<P>$  \\ \hline
  0.5            &    500           & $0.0107 \pm 0.0021$ &  7.9             \\
  0.5            &   1000           & $0.0094 \pm 0.0018$ &  9.1             \\
  1.0            &    500           & $0.0127 \pm 0.0024$ &  6.7             \\
  1.0            &   1000           & $0.0150 \pm 0.0028$ &  5.7             \\ \hline
\end{tabular}
\end{center}
\end{table}

We calculate the mean 1D power spectrum at $k=0.049 h {\rm Mpc}^{-1} $ 
for each survey geometry, using the real space three dimensional power 
spectrum recovered in each of four zones into which the 
APM Survey was split by Baugh \& Efstathiou (1993), with the 
results given in Table \ref{tab:mpr}.
For geometry B, we compute the mean expected power for two 
apparent magnitude limits.
In case C for multiple pencil beams, we use two estimates of the 
real space power spectrum in 3D.
The parameter $\alpha$ describes the evolution of clustering 
with redshift; for $\alpha=0$ clustering is fixed in 
comoving co-ordinates, for $\alpha=2$ clustering evolves according to 
linear perturbation theory.

In the case of a uniform cylinder, we note that the APM power spectrum 
gives a mean power that is lower than that obtained with a 
$ \xi(r) = (r/5)^{1.8} $ power law extended to all scales by $ 30 \%$ (KP 
find $\langle P \rangle = 0.019$).
Furthermore, if we truncate the APM power spectrum at $\lambda = 30 \mpc$, 
the mean power for a cylinder falls by a factor of two.
More realistic modelling of the survey window function boosts the 
mean power by a factor of $3-5$, confirming the claim of KP that there 
is little evidence for periodicity in the deep NGP + SGP beams alone.

The final column in Table \ref{tab:mpr} gives the ratio of the observed 
peak power to the external estimate of the mean power.
For the BEKS data, (geometries A and B), the peak power is 
$ P = 80.92 $.
KP point out that 380 galaxies have redshifts less than $z=0.5$, 
which gives an amplitude in our units of $ P = 80.92/380 = 0.213$,
which is used to calculate the peak to mean power ratio for cases A and B.
Figure 4 of Broadhurst \and (1995) gives the height of the peak 
for many beams as $P=0.085$, which is used in case C.

KP found that projected power spectra have an exponential 
distribtuion of amplitudes, as expected from the central limit 
theorem (Fan \& Bardeen 1995): the probability of observing a 
peak in the power spectrum in excess of $P$ at some 
selected frequency given the mean power $ \langle 
P \rangle$ is (Kendall \& Stuart 1977) $ Pr(> P) = \exp( -P / 
\langle P \rangle ) $.

There is a factor of $\sim 2$ difference in the peak to mean power 
ratio for different survey geometries which illustrates the 
difficulty in making an external estimate of the mean power.
When the effects of redshift space distortions are included, the 
peak to mean power ratio can fall as low as 5.7, which gives 
a field point probability that the peak arises from a Gaussian 
random field of $3.3 \times 10^{-3}$. 
This is much higher than the most pessimistic example in which 
there are no redshift space distortions and $\alpha=0$, where 
the probability is $4.5 \times 10^{-5}$.
However, in both cases the field point probability is much larger 
than the figure $4 \times 10^{-7}$ quoted by Broadhurst \and (1995) 
for the null hypothesis of no large scale clustering of galaxies.
Note that the amplitude of the APM power spectrum could be low by 
as much as $20 \%$, due to merging corrections to APM images which 
we have not applied here (Maddox \and 1990, 1996) or uncertainties 
in the form of the redshift distribution (Baugh \& Efstathiou 1993, 
Gazta\~{n}aga 1995).

We have simulated 1D power spectra using the projected amplitude 
for $P_{1D}(k)$ computed from the APM power spectrum after convolution 
with the multiple beam window function.
Real and imaginary Fourier components are generated with a Gaussian 
distribution.
The resolution of the 1D power spectrum is taken to be 
$\delta k = 0.007 {\rm h Mpc}^{-1}$ as calculated by KP for 
the combined sample of BEKS and which matches the full width at half 
maximum of the peak in Figure 4 of Broadhurst \and (1995).
Examples of these synthetic spectra are shown in Figure \ref{fig:pk}.
We generate a large number of realisations of the Gaussian power spectrum 
and look for peaks in the wavenumber range $0.028 {\rm h Mpc}^{-1} \le k
\le 0.063 h {\rm Mpc}^{-1}$, corresponding to wavelengths of 
$ 100 h^{-1} {\rm Mpc} < \lambda < 224 h^{-1} {\rm Mpc}$.  
We find peaks in excess of the height reported by Broadhurst \and (1995) 
$3\%$ of the time using the estimated real space mean power, rising to 
$6\%$ of realisations when redshift space distortions with $\beta= 1.0$ and 
$ \sigma_{v} = 500 {\rm kms}^{-1}$ are included.
A typical power spectrum generated with a high peak in the range reported by 
Broadhurst \and is shown in Figure \ref{fig:pk}(b). 
The power spectrum realisation contains several peaks that are a few  
times higher than the mean power, indicated by the dotted line, in contrast 
with the power spectrum shown by Figure 4 of Broadhurst \and.
Roughly 1 in every 1500 realisations contains a peak in excess of $P = 0.085$ in the 
specified wavenumber range,  without having another peak in the range $0.0 < k < 0.3$ that 
is higher than $P = 0.02$; two examples are shown in Figure \ref{fig:pk}(a) and (c).

\section{Discussion}

We have examined the compatibility of the three dimensional 
power spectrum measured from the APM Galaxy Survey 
(Baugh \& Efstathiou 1993, 1994) with the power spectrum 
recovered from deep pencil beam surveys by 
BEKS, Szalay \and (1993) and Broadhurst \and (1995).
The probability of observing a peak in the 1D power spectrum, 
given our knowledge of the 3D power spectrum on these scales 
depends upon the ratio of the height of the observed 
peak compared with the external estimate of the mean power.
External estimates of the mean power in the 1D datasets 
are extremely sensitive to how the survey geometry is modelled  
and to the magnitude of the redshift space distortions.
For a survey consisting of many pencil beams, we find 
a factor of two difference in the amplitude of the mean 
power that we calculate, for various models of the 
distortions in the pattern of clustering due to the 
peculiar velocities of clusters.

In the case of our highest estimate of the mean power 
for a survey consisting of multiple pencil beams, 
we find that the field point probability of finding a 
peak of the observed height given the null hypothesis 
of the measured clustering in 3D, is roughly 1 in 300.
We have also generated Gaussian realisations of the pencil beam 
power spectra, using the 1D power spectrum obtained by 
convolving the measured 3D power spectrum with the survey 
geometry.
Using the same resolution as the observed 1D power spectrum, 
we find that peaks of the observed height or higher in 
the wavelength range $100 \le \lambda \le 200 h^{-1} {\rm Mpc}$ 
occur in $3-6 \%$ of the realisations, depending upon the 
model  assumed for the redshift space distortions.

The general appearence of these synthetic power spectra is quite 
different from that of the power spectrum of 
the observations in the majority of realisations.
We find that a high peak in the power in the wavelength 
range specified above 
is generally accompanied by peaks that are a few times the amplitude of 
the mean power at other wavenumbers.
In only a small number of cases, roughly 1 in 3000 do with find 
an isolated high peak with the other peaks in the spectrum less than 
twice the amplitude of the mean.  
KP discuss the effects that redshift errors or binning of the 
data in real space before taking the Fourier transform could have 
upon the appearence of the observed power spectrum.
Also, we have neglected the consequences of the density distribution 
in 3D being skewed ({\it e.g.} Efstathiou \and 1990, 
Saunders \and 1991, Gazta\~{n}aga 1994) which has been shown to further 
enhance the probability of observing a peak given a 
null hypothesis of the measured 3D clustering (Amendola 1994).

The ability of the measured 3D clustering to account for the presence of
high peaks in the observed 1D power spectrum makes it seem unlikely
that large scale structure can be responsible for the steep observed slope
of the galaxy counts at bright magnitudes, a conclusion already
discussed by Maddox \and (1990) and Loveday \and (1992).
The factor of two increase in the normalisation of the luminosity function
desired to provide a better match to the faint counts (Shanks 1990) 
requires that the local universe out to a radius of
around $130 h^{-1} {\rm Mpc}$,
corresponding to the median redshift of galaxies brighter
than $b_{J} = 17$, is massively underdense.
The variance in the number of galaxies
measured on these scales from the APM Galaxy Survey is
on the order of $10 \%$ (Baugh \& Efstathiou 1993, Gazta\~{n}aga 1994)
many times smaller than would be necessary to result in a factor of
two fluctuation in the number of galaxies found in a volume of such
a large radius (see also the discussion in Glazebrook
\and 1994).

\bigskip

\section*{Acknowledgements}
The author acknowledges George Efstathiou, Enrique Gazta\~{n}aga 
and Shaun Cole for useful conversations and comments on an earlier 
version of the manuscript.
 
\bigskip

\setlength{\parindent}{0mm}
{\bf REFERENCES} 
\bigskip

\def\refe {\par \hangindent=.7cm \hangafter=1 \noindent}
\def\aj { ApJ, }
\def\mn { MNRAS, }
\def\apl { Ap. J. (Letters), }

\refe Amendola, L., 1994, \aj 430, L9
\refe Baugh, C.M., Efstathiou, G., 1993, \mn 265, 145
\refe Baugh, C.M., Efstathiou, G., 1994, \mn 267, 323
\refe Baugh, C.M., 1996, \mn 280, 267
\refe Broadhurst, T.J., Ellis, R.S., Shanks, T., 1988, \mn 235, 827
\refe Broadhurst, T.J., Ellis, R.S., Koo, D.C., Szalay, A.S., 1990, 
      Nature, 343, 726 (BEKS) 
\refe Broadhurst, T.J., Szalay, A.S., Ellis, R.S., Ellman, N., Koo, D., 
      1995, in Wide Field Spectroscopy in the Distant Universe, ed. 
      Maddox, S.J., Arag\'{o}n-Salamanca, World Scientific
\refe Cole, S., Fisher, K.B., Weinberg, D.H., 1995, \mn 275, 515
\refe Colless, M., Ellis, R.S., Taylor, K., Hook, R.N., 1990, \mn 
      244, 408
\refe Colless, M., Ellis, R.S., Broadhurst, T.J., Taylor, K., Peterson, B.A., 
      1993, \mn 261, 19
\refe Davis, M., Peebles, P.J.E., 1983, \aj 267, 465
\refe Efstathiou, G., Kaiser, N., Saunders, W., Lawrence, A., 
      Rowan-Robinson, M., Ellis, R.S., Frenk, C.S., 1990, \mn 
      247, 10p
\refe Fan, Z.H., Bardeen, J.M., 1995, Phys. Rev. D., 51, 6714
\refe Feldman, H.A., Kaiser, N., Peacock, J.A., \aj 426, 23
\refe Gazta\~{n}aga, E., 1994 \mn 268, 913
\refe Gazta\~{n}aga, E., 1995 \aj 454, 561
\refe Glazebrook, K., Peacock, J.A., Collins, C.A., Miller, L., 1994, \mn
      266, 65
\refe Glazebrook, K., Ellis, R., Santiago, B., Griffiths, R., 1995, \mn 
      275, L19
\refe Kaiser, N., 1987, \mn 227, 1
\refe Kaiser, N., Peacock, J.A., 1991, \aj 379, 482 (KP)
\refe Kendall, M., Stuart, A., 1977 Advanced Theory of Statistics, 
      Macmillan
\refe Landy, S.D., Shectman, S.A., Lin, H., Kirshner, R.P., Oemler, A.A., 
        Tucker, D., 1996, \aj 456 L1
\refe Loveday, J., Peterson, B.A., Efstathiou, G., Maddox, S.J., 1992, 
      \aj 390, 338
\refe Luo, A., Vishniac, E.T., 1993, \aj 415, 450
\refe Maddox, S.J., Efstathiou, G., Sutherland, W.J., Loveday, J., 1990, 
      \mn 242, 43p
\refe Park, C., Vogeley, M.S., Geller, M.J., Huchra, J.P., 1994, 
      \aj 431, 569
\refe Peacock, J.A., 1991, \mn 253, 1p
\refe Peacock, J.A., Dodds, S.J., 1994, \mn 267, 1020
\refe Saunders, W., Frenk, C.S., Rowan-Robinson, M., Efstathiou, G., 
      Lawrence, A., Kaiser, N., Ellis, R.S., Crawford, J., Xia, X.Y., 
      Parry, I, 1991, Nature, 349, 32 
\refe Shanks, T., 1990, in {\it The Galactic and Extragalactic Background 
      Radiations}, 269, eds Bowyer, S., Leinhert, C, Kliwer, Dordrecht
\refe Szalay, A.S., Broadhurst, T.J., Ellman, N., Koo, D.C., Ellis, R.S., 
      1993, Proc. Nat. Acad. Sci., 90, 4853
\refe Tadros, H., Efstathiou, G., 1996, \mn submitted

\end{document}